# Gravity and motion

A. LOINGER

Dipartimento di Fisica, Università di Milano

Via Celoria, 16 − 20133 Milano, Italy

**Summary.** − Several arguments concerning the relativistic *vexatae quaestiones* of the gravity field of a point mass and of the wavy gravity fields.

## Introduction

I give a concise analysis, with essential historical references, of two critical subjects of relativistic astrophysics: the gravity field of a point mass and the wavy gravity fields.

## First Part: On the gravity field of a point mass

**1**. − The solution of the problem of the Einsteinian gravitational field, which is generated by a point mass $M$ at rest, is given − if $r,\theta,\varphi$ are spherical polar coordinates − by the following expression of the spacetime interval [1], [1bis]:

$$(1.1) \qquad ds^2 = \left[1 - \frac{2m}{f(r)}\right] c^2 dt^2 - \left[1 - \frac{2m}{f(r)}\right]^{-1} [df(r)]^2 - f^2(r)\, d\omega^2 \;,$$

where: $m \equiv GM/c^2$; $G$ is the gravitational constant and $c$ the speed of light *in vacuo*; $d\omega^2 \equiv d\theta^2 + \sin^2\theta\, d\varphi^2$; $f(r)$ is any regular function of $r$.

In the *original* solution form of the above problem given by Schwarzschild in 1916 [2], the function $f(r)$ is as follows:

$$(1.2) \qquad f(r) \equiv \left[ r^3 + (2m)^3 \right]^{1/3} \;;$$

thus Schwarzschild's $ds^2$ holds, physically and mathematically, in the *entire* spacetime, with the only exception of the origin $r = 0$, seat of the mass $M$, where we have a singularity.

If one chooses simply

$$(1.3) \qquad f(r) \equiv r \;,$$

one obtains the so-called *standard* form of solution, which is usually, *and erroneously*, named "by Schwarzschild". It was deduced *ex novo*, by integrating the Einstein equations, by Hilbert [3], by Droste [4], and by Weyl [5], independently.

Another interesting form, first investigated by M. Brillouin [6], is obtained by putting in (1.1)

$$(1.4) \qquad f(r) \equiv r + 2m \;;$$

it holds in the *whole* spacetime, with the only exception of the origin.

On the contrary, the *standard form is physically valid* **only for** *r>2m*, because within the spatial surface *r=2m* (which is a singular locus) the time coordinate takes the role of the radial coordinate, and *vice versa*, the solution becomes *non*-static, and the d$s^2$ loses its essential property of physical "appropriateness", according to the expressive Hilbert's terminology [3]. Further, I emphasize with Nathan Rosen that the radial coordinate of the standard solution has been initially chosen in such a way that the area of spatial surface *r=k* is given by $4\pi k^2$. Accordingly, it is difficult to admit that the coordinate *r* can transform itself into a time coordinate. We ask ourselves: does the restriction *r>2m* imply a *physical* limitation? Not at all! Indeed, as the classic Authors knew, the exterior part *r>2m* of the standard form is *diffeomorphic* to the Schwarzschild's and Brillouin's forms, which hold for *r>0*. One can say that the "*globe*" *r=2m* of the standard form shrinks into the *point r=0* of Schwarzschild's and Brillouin's forms, which is a singular point with an associate superficial area $4\pi(2m)^2$.

An odd reflection on the "*globe*" *r=2m* generated the notion of black hole: it would not have come forth if the treatises had expounded the forms of Schwarzschild or of Brillouin, in lieu of the standard form.

In a review article on the black holes [7], a true manifesto of scientific policy, we find some amazing results, e.g. the following evaluation of the average density of a black hole (*sic*): mass *M* divided by $(4/3)\pi r_0^3$, where $r_0 \equiv 2GM/c^2$; thus the *point* mass *M* is ideally distributed within the "globe" *r=2m*; accordingly, the average density of a black hole is inversely proportional to the square of its mass *M*. A marvellous consequence: if *M* is equal to the mass $M_\odot$ of the Sun, the average density is $\approx 10^{16} \text{g} \cdot \text{cm}^{-3}$, whereas for a mass $M = 10^8 M_\odot$, the average density is $\approx 1 \text{ g} \cdot \text{cm}^{-3}$, i.e. equal to water density. And a black hole of vanishing density has an infinitely large mass, and *vice versa*.

I remark that as far back as 1922 the competent scientists knew the right interpretation of the standard solution. Indeed, in 1922 a meeting was held at Collège de France, which was also attended by Einstein; the physical meaning of the "*globe*" *r=2m* was discussed and definitively clarified. See the lucid paper by Marcel Brillouin quoted in [6]. This author shows that it is not permitted to extend the radial coordinate of Schwarzschild's and Brillouin's forms to the negative values of the interval −2m<r<0, and proves simultaneously that the attribution of a physical meaning to the interval 0<r<2m of the standard form is pure nonsense. Furthermore, let us recall that in a second fundamental memoir [8], Schwarzschild determined the Einsteinian gravitational field generated by an incompressible fluid sphere; now, if one computes the limit of his solution when the sphere contracts into a material point of a finite mass *M*, one finds anew the Schwarzschildian solution for a point mass: this is another proof of the "physicality" of the origin. Moreover, we remember that a fluid sphere of uniform density and *given mass* cannot have a radius smaller than (9/8)(2m).

Quite similar considerations can be made for the gravitational fields generated by an electrically charged particle and by the spinning particle of the well-known Kerr's solution. In regard to the solution form − *non*-static and "maximally extended" − of Schwarzschild problem due to Kruskal [9] and Szekeres [10], we can declare its *physical superfluity*, because already the static forms of Schwarzschild and Brillouin, *in particular*, are "maximally extended".

Question concerning the *continued gravitational collapse*: it is almost evident that if we bear in mind, e.g., Schwarzschild's and Brillouin's forms of solution, no continued collapse can generate a black hole − and this was just Einstein's opinion. (See also APPENDIX A). On the other hand, it is physically clear that the real gravitational collapses cannot continue indefinitely, but end finally in astronomical objects of finite, relatively small, dimensions. (See my article "Relativistic spherical symmetries", http://xxx.lanl.gov/abs/physics/0107071 (July 28th, 2001) (misclassified: proper class.: gr-qc)).

I remark at last that the most cautious among the *observational* astrophysicists have always called in question the very notion of black hole. They know perfectly that the *observed* "black holes" do not coincide with the theoretical black holes, but are only large, or enormously large, masses concentrated in relatively small volumes.

**1bis**. − It can be shown (see e.g. P.A.M. Dirac, *General Theory of Relativity* − Wiley and Sons, New York, *etc.*, 1975 − p.32 and foll.) that the singular surface $r = 2m$ of the *standard* form of solution to Schwarzschild problem has the following properties:

*i*) A material point falling into the central body takes an infinite time to reach the critical surface $r = 2m$;

*ii*) If the above particle is emitting light of a certain frequency and is being observed by someone at a *large* value of *r*, its light is red-shifted by a factor $(1 - 2m/r)^{-1/2}$. This factor becomes infinite as the particle approaches the singular surface $r = 2m$. All the physical processes *on* the particle will be observed to be going more and more slowly as it approaches $r = 2m$;

*iii*) Let us consider an observer travelling *with* the particle; it reaches $r = 2m$ after the lapse of a *finite proper time* for the observer, who has aged only a finite amount when he and the particle reach $r = 2m$;

*iv*) The spatial region $r < 2m$ cannot communicate with the space for which $r > 2m$. Any signal, even a light signal, would take an infinite time to cross the boundary $r = 2m$; thus we cannot have a direct observational knowledge of the region $r < 2m$. Such a region is called a *black hole*, because things may fall into it (taking an infinite time by our clocks, to do so), but nothing can come out. −

I have reproduced almost literally some significative sentences of sect.**19** of the cited Dirac's booklet. The proofs of properties *i*), *ii*), *iii*) are quite rigorous. On the contrary, the proof of property *iv*) rests on a paralogism: indeed, if one *hides* (as Dirac and many authors do) the singularity $r = 2m$ in the connection between the coordinates $(r,t)$ and *suitable* new coordinates $(\rho, \tau)$, it is possible to extend the *transformed* solution to the region $r < 2m$ (where the roles of *r* and *t* are interchanged!). Then, one can "prove" formally property *iv*).

I remark finally that properties *i*), *ii*), *iii*) hold also for Schwarzschild's and Brillouin's forms (e.g.) if we substitute the critical surface $r = 2m$ of the standard form with the point $r = 0$ of the above forms.

In lieu of property *iv*), which characterizes the odd notion of black hole, the mentioned forms have the following property: any signal which starts from $r = 0$ will take an infinite time to reach any finite distance from the origin: it seems that one must be content with a black *point*. However, the story is not concluded, because it is possible to show that there exist infinite, non-trivial forms of solution to Schwarzschild problem which are *regular everywhere for* $r \geq 0$ (see my article "Regular solutions of Schwarzschild problem", http://xxx.lanl.gov/abs/physics/0104064 (April 20th, 2001) (misclassified: proper class.: gr-qc))).

**2**. – Many physicists think that the notion of black hole is only a relativistic generalization of a Newtonian notion, created by Michell (1784) and Laplace (1796). Now, as it has been proved by McVittie (see *The Observatory*, **78** (1978) 272), this idea is based only on "a play of words in expressions such as *the velocity of escape* or *the escape from a body*".

Let us indeed consider a celestial spherical body of radius $R$ and mass $M$. According to Newtonian dynamics, the velocity of escape $w$ of a particle, which is projected radially outwards from the body's surface, is given by

$$(2.1) \qquad w^2 = 2GM/R \quad ,$$

If the particle is projected with a velocity $u$ smaller than $w$, i.e. if

$$(2.2) \qquad u^2 < 2GM/R \quad ,$$

it will arrive at a finite distance from the celestial body, and then will fall on its surface again.

By employing the Newtonian corpuscular theory of light, which says that light is composed of corpuscles obedient to Newton's law of gravitation and travelling with a given velocity $c$, Michell and Laplace remarked that if $c < w$ the light corpuscles cannot go away indefinitely from the celestial body. Only if the radius $R$ of the celestial body is such that

$$(2.3) \qquad R = 2GM/c^2 \quad ,$$

they can escape from the gravitational attraction exerted by the mass $M$. Then, if

$$(2.4) \qquad R < 2GM/c^2 \quad ,$$

the light corpuscles will attain to a finite distance $D$ from the celestial body, and an observer situated at an intermediate distance between $R$ and $D$ will see the celestial body, owing to the light corpuscles which arrive at his eyes.

If only the spherosymmetrical black hole of general relativity existed, it ought to have the fundamental property that neither the material particles nor the light corpuscles can leave its surface (see sect.**1bis**). Therefore such an object would be invisible to any observer, however near he may be. None of the phenomena observed by the experimentalists in the region surrounding a "black hole" of Michell-Laplace is present in the neighbourhood of a black hole of general relativity.

The imagined connexion with the Newtonian formula (2.3) comes forth in this way: if for the determination of the Einsteinian gravitational field generated by a point mass $M$ (Schwarzschild problem) one chooses the *standard* form – see (1.3) – as Droste, Hilbert and Weyl (*not Schwarzschild*) did, the radial coordinate $r_0$ of the points of the space surface $r = r_0$ ($r_0$ is the "radius of the black hole") is given by

$$(2.5) \qquad r_0 \equiv 2GM/c^2 \quad ;$$

this formula resembles the Newtonian formula (2.3), which regards a velocity of escape $c$. But eq.(2.3) implies obviously that all observers – including those at an infinitely great distance from the celestial body – can see it. As it is clear, the "Newtonian black hole" of Michell-Laplace is *not* a black hole!

The well-known scepticism of McVittie regarding the existence of relativistic black holes appears clearly from the final considerations of the cited Note, where he emphasizes, in particular, that there is "no way of asserting through some analogy with Newtonian gravitational theory that a black hole could be a component of a close binary system or that two black holes could collide. An existence theorem would first be needed to show that Einstein's field equations contained solutions which described such configurations".

**3**. − A question: why did the standard form (Droste-Hilbert-Weyl) prevail − in the scientific literature − over the original form of Schwarzschild? This is an interesting problem for an unprejudiced historian of our science. Here I limit myself to mention three reasons: *i*) the mathematical deduction of the standard form is simpler than that of Schwarzschild's form; *ii*) the influence implicitly exerted by Hilbert, the greatest mathematician and mathematical physicist of past century; *iii*) the premature death of Schwarzschild, due to a rare illness contracted at the German-Russian front.

**4**. − See in [23] a complete list of my papers concerning the subject of previous sects. **1**÷**3** published in Los Alamos Archive.

**Second Part: On the wavy gravity fields**

**5**. − During an epistolary discussion a known relativist wrote to me: "Without gravity waves, one would have to explain an instantaneous propagation of a change in the metric over the whole universe simply by changing the distribution of stress or mass in a system". This conviction is quite widespread, but is wrong. It is an incontestable *fact* that the physical non-existence of the gravitational waves is quite consistent with the fundamental principles of relativity theory: the Einsteinian field equations are time-symmetrical − and therefore it is perfectly legitimate to discard formal solutions which are time-asymmetrical. Analogously, Maxwell equations of the electromagnetic field are time-symmetrical: the existence of the electromagnetic waves is only a *theoretical possibility*, **not** *a theoretical necessity*: the *physical* existence of the e.m. waves is an **experimental** fact.

**6**. − *A question*: what is the behaviour of the metric tensor when, e.g., a *supernova* explodes? *Answer*: A sufficiently near apparatus would register a variation of the Einsteinian gravitational field which would be approximately similar to the corresponding variation of the Newtonian field. However, no gravitational *wave* − i.e. no physical entity endowed with a "life" *independent* of the source − would be emitted.

**7**. − There is an enormous number of papers concerning the gravitational waves. Here I limit myself to quote two manifestos by Schutz [11]; the first of them includes an ample bibliography.

Originally, the emission of gravitational waves was hypothesized and calculated *in analogy to the electromagnetic case*, starting from the **linear** approximation of the Einsteinian field equations, whose spatio-temporal substrate is "rigid" and coincides simply with Minkowski's spacetime. This "rigidity" tells us that there is a *primary conceptual difference* between exact theory and linearized theory, which is simply a theory of a *weak* gravitational field in a *flat* spacetime, having an invariant character with respect to the Lorentz transformations. Its formalism resembles the Minkowskian formalism of Maxwell theory, and − under suitable boundary conditions − allows the theoretical emission of undulating fields. Einstein did not like this result − and for many reasons. In spite of the innumerable computations that were performed since the Twenties of past century, he doubted always about the physical reality of the gravitational radiation. In particular, Einstein thought it is likely that only the *time-symmetrical* solutions of his field equations can represent physical phenomena.

I remark that the usual computations concerning the emission of gravitational waves by moving bodies are of a *perturbative* character and have the linearized version of the theory as a first approximation. Accordingly, they do not yield a true **existence theorem**. On the other hand, it is possible to prove *rigorously* that no motion of point masses can generate gravitational waves [12]. A very simple proof is the following. Let us suppose that at a given instant $t$ of its motion a given point mass $M$ begins to send forth a gravitational wave and let us assume to know the *kinematical characteristics* of the motion between $t$ and $t + |dt|$. It is indisputabile that we can reproduce *these* characteristics in a gravitational motion of mass $M$ in a suitable "external" gravitational field, within a time interval equal to $|dt|$, conveniently chosen. But in this case the mass $M$ moves along a **geodesic** − and therefore it cannot emit any gravitational radiation: indeed, the geodesic motions are "free" motions, they are the perfect analogues of the rectilinear and uniform motions of an electric charge of the usual Maxwell-Lorentz theory. *Q.e.d.*

Conclusion: since *no "mechanism" exists for the generation of gravitational waves* (the restriction to motions of mass points is conceptually inessential), all the formal solutions of the Einsteinian field equations having an undulatory character do *not* describe *physical* waves. (See also APPENDIX $B_1$).

There are, however, other arguments which demonstrate the physical non-existence of the gravity radiation. Consider, for instance, that in the *exact* theory a gravitational wave would be an entity destitute of a *true* energy and a *true* momentum: consequently it cannot interact with any whatever apparatus or with an e.m. field: otherwise the energy-momentum account would not balance.

Many years ago, in the end of a paper [13] Pirani proposed to the reader and to himself the following problem: "Suppose for example that a Schwarzschild particle is disturbed from static spherical symmetry by an internal agency, radiates for some time, and finally is restored to static spherical symmetry. Is its total mass necessarily the same as before?" We have here a typical *Scheinproblem*: if the gravitational radiation existed, it would have only a pseudo (false) energy, therefore the final mass would be identical to the initial mass.

Furthermore, the undulatory character and the propagation velocity of a metric tensor *depend* on the reference system: with a suitable choice of the frame the undulatory character disappears, with a suitable choice of the frame the propagation velocity can take any value between zero and the infinite. (In general relativity we do *not* have a class of physically privileged frames of reference …).

**8**. − Several authors have avoided intentionally the basic problem concerning the emission "mechanism" of the gravity waves and have looked for undulating solutions of the Einsteinian equations with a mass tensor equal to zero. Some exact solutions and others of a perturbative nature have been found. This is not surprising because the theory of the characteristics of the Einstein equations (Levi-Civita [14]) yields a rigorous proof of the existence of *wave fronts*; of what *kind* of waves? *Electromagnetic* waves, according to Levi-Civita − for several reasons, *in primis* because general relativity (analogously to special theory) must contain the geometrical optics.

Einstein, Møller, Scheidegger [15] and Rosen [16], but particularly Infeld e Plebanski [17] had serious reasons against the physical existence of the gravity waves, see APPENDIX $B_2$.

**9**. − According to a diffuse belief, the time decrease of the revolution period of the famous binary radiopulsar PSR1913+16 gives an experimental *indirect* proof of the physical reality of the gravitational radiation.

Owing to the observational data yielded by the "regular clock" of the pulsar, the interesting orbital parameters and the masses of the two stars (regarded as point objects) have been perturbatively computed. Then, the *perturbative* quadrupole formula gave a decrease of the revolution period, which agreed very well with the observations.

I emphasize the following points. In the *exact* theory the quadrupole formula loses any meaning because the hypothesized gravity waves do *not* have a *true* energy. Therefore, the *true* mechanical energy which is lost during the revolution motion ought to transform itself into the *pseudo* energy of the hypothetical gravity radiation: evidently the energy account does *not* balance.

Devil's advocate could object: if we restrict ourselves to the *linear* approximation of general relativity (as the experimentalists do), which has Minkowski spacetime as its substrate, the physical existence of the gravitational waves is surely a theoretical possibility. *Answer*: the energy-momentum of such gravitational waves has a tensor character only under Lorentz transformations, not under general transformations. Therefore it is always possible to find (and we remain, of course, in the ambit of the linear approximation) a general frame for which the above energy-momentum is equal to *zero*. But a wave with no energy and no momentum is not a physical object, even if it is formally endowed with a curvature tensor different from zero.

In the second place, there are realistic explanations of the decrease of the revolution period − as it is well known to the observational astrophysicists; for instance, viscous losses of the pulsar companion would give a time decrease of the revolution period of the same order of magnitude of that given by the hypothesized emission of gravity radiation.

Finally, the empirical success of a theory − or of a given computation − is not an *absolute* guaranty for its *conceptual* adequacy. Consider for instance the Ptolemaic theory of cycles and epicycles, which explained rather well the planetary orbits (with the only exception of Mercury's). As it was emphasized by Truesdell [19], the heliocentric theory would have been rejected if people of 17th century had had the modern computers.

**10**. − See in [24] a complete list of my papers concerning the subject of previous sects. **5÷9** published in Los Alamos Archive.

> "Ti par che farrebe male un che volesse
> mettere sotto sopra il mondo rinversato?"
> Giordano Bruno

**APPENDIX A**

All the Great Spirits who created and developed the general relativity (Einstein, Levi-Civita, Schwarzschild, Hilbert, Weyl, Eddington, Pauli, Fock, …) rejected always the very notion of black hole. In 1939 Einstein wrote a remarkable article [19], which was efficaciously summarized by Bergmann [20] with the following sentences, where the phrase "Schwarzschild singularity" means *more solito* (and improperly!) the critical surface $r = 2m$ of the *standard* form of solution to Schwarzschild problem. "Einstein investigated the field of a system of many mass points, each of which is moving along a circular path, $r = $ const., under the influence of the field created by the ensemble. If the axes of the circular paths are assumed to be oriented at random, the whole system or cluster is spherically symmetric. The purpose of the investigation was to find out whether the constituent particles can be concentrated toward the center so strongly that the total field exhibits a Schwarzschild singularity. The investigation showed that even before the critical concentration of particles is reached, some of the particles (those on the outside) begin to move with the velocity of light, that is, along zero world lines. It is, therefore, impossible to concentrate the particles of the cluster to such a degree that the field has a singularity. (The singularities connected with each individual mass point are, of course, not considered.)

Einstein chose this example so that he would not have to consider thermodynamical questions, or to introduce a pressure, for the particles of his cluster do not undergo collisions, and their individual paths are explicitly known. In this respect, Einstein's cluster has properties which are nowhere encountered in nature. Nevertheless, it appears reasonable to believe that Einstein's result can be extended to conglomerations of particles where the motions of the individual particles are not artificially restricted as in Einstein's example." [20].

(*N.B.* − In reality, Einstein [19] employed the so-called isotropic coordinates in lieu of the standard coordinates. Of course, the validity of his argument is independent of this choice.).

**APPENDIX B$_1$**

*i*) Any particle of a continuous, incoherent "cloud of dust", characterized by the mass tensor

(B$_1$.1) $$T^{jk} = \rho \frac{\mathrm{d}x^j}{\mathrm{d}s} \frac{\mathrm{d}x^k}{\mathrm{d}s} , \quad (j,k = 0,1,2,3) ,$$

where $\rho$ is the invariant mass density, describes a *geodesic* line, and therefore cannot emit gravitational waves (see the first paper quoted in [12]). A simple application: the gravitational motions of the members of solar system.

*ii*) A well-known Fermi's geometrical theorem [21] as generalized by Eisenhart [22] affirms: For a manifold endowed with a *symmetric* connection it is possible to choose a coordinate system with respect to which the components $\Gamma^i_{jk}$ ($= \Gamma^i_{kj}$) of the connection are *zero* at all points of a curve (or of a portion of it).

For a Riemann-Einstein spacetime this means that there exists a coordinate system with respect to which the first derivatives of the components $h_{jk}$, ($j,k = 0,1,2,3$), of the metric tensor are *zero* at all points of a curve (or of a portion of it) − in particular, at all points of a time-like world line.

*iii*) Let us now consider a continuous medium (for instance, a perfect fluid) characterized by a certain mass tensor $T_{jk}$, and let $g_{jk}(x)$ be the solutions of Einstein equations

(B$_1$.2) $$R_{jk} - \frac{1}{2} g_{jk} R = -\kappa T_{jk}$$

corresponding to a generic motion of our medium with respect to a given reference frame $(x) \equiv (x^0, x^1, x^2, x^3)$. Let us suppose to follow the motion of a given mass element describing a certain world line *L*. If we refer this motion, from the initial time $t_0$ on, to a Fermi's reference system $(z) \equiv (z^0, z^1, z^2, z^3)$, the components $h_{jk}(z)$ of the metric tensor will be equal to some constants for *all* points of line *L*. In other words, the gravitational field *on L* has been obliterated. Consequently, no gravitational wave has been sent forth. Now, line *L* is quite generic, and therefore *no motion of the continuous medium can give origin to a gravitational radiation*.

**APPENDIX B$_2$**

By means of approximation methods for the treatment of gravitational motions of the bodies Scheidegger in 1953 [15], and Infeld and Plebanski in 1960 [17] arrived at negative conclusions about the *physical* existence of a gravity radiation.

Scheidegger showed that all the computed radiation terms can be destroyed by suitable coordinate transformations. Infeld and Plebanski showed that "…it is hardly possible to connect any physical meaning with the flux of energy and momentum tensor defined with the help of the pseudo-energy-momentum tensor. Indeed, the radiation can be annihilated by a proper choice of the coordinate system. On the other hand, if we use a coordinate system in which the flux of energy may exist, then it can be made whatever we like by the addition of proper harmonic functions…".

The common conclusion of the arguments of Appendices B$_1$ and B$_2$, in particular, is that there is no "mechanism" apt to produce gravitational waves. A conclusion which is in full accord with Einstein's ideas and Levi-Civita's conviction.